# A feasibility study on SSVEP-based interaction with motivating and immersive virtual and augmented reality

J Faller[1], B Z Allison[1], C Brunner[1], R Scherer[1], D Schmalstieg[2], G Pfurtscheller[1] and C Neuper[1,3]

[1]Institute for Knowledge Discovery, Laboratory of Brain-Computer Interfaces, Graz University of Technology, Krenngasse 37, 8010 Graz, Austria
[2]Institute for Computer Graphics and Vision, Graz University of Technology, Inffeldgasse 16c, 8010 Graz, Austria
[3]Department of Psychology, University of Graz, Universitätsplatz 2/III, 8010 Graz, Austria

E-mail: josef.faller@tugraz.at

**Abstract**

Non-invasive steady-state visual evoked potential (SSVEP) based brain-computer interface (BCI) systems offer high bandwidth compared to other BCI types and require only minimal calibration and training. Virtual reality (VR) has been already validated as effective, safe, affordable and motivating feedback modality for BCI experiments. Augmented reality (AR) enhances the physical world by superimposing informative, context sensitive, computer generated content. In the context of BCI, AR can be used as a friendlier and more intuitive real-world user interface, thereby facilitating a more seamless and goal directed interaction. This can improve practicality and usability of BCI systems and may help to compensate for their low bandwidth. In this feasibility study, three healthy participants had to finish a complex navigation task in immersive VR and AR conditions using an online SSVEP BCI. Two out of three subjects were successful in all conditions. To our knowledge, this is the first work to present an SSVEP BCI that operates using target stimuli integrated in immersive VR and AR (head-mounted display and camera). This research direction can benefit patients by introducing more intuitive and effective real-world interaction (e.g. smart home control). It may also be relevant for user groups that require or benefit from hands free operation (e.g. due to temporary situational disability).

# 1. Introduction

Brain-computer interface (BCI) systems (Wolpaw et al. 2002, Pfurtscheller et al. 2006) establish an additional, direct channel of communication and/or control between the human brain and a computer. BCIs try to determine user intention based on the classification of characteristic spatial, temporal or spectral features (i.e. intentional, goal-directed mental activity) extracted from a brain signal like the electroencephalogram (EEG) and provide feedback in real-time.

Any visual stimulus that oscillates at a fixed frequency above 6 Hz generates steady-state visual evoked potentials (SSVEP) over the occipital cortex. The user can modulate these SSVEP by focusing attention on one of multiple stimuli. SSVEP-based BCIs (Cheng et al. 2002, Gao et al. 2003) are widely used, since the majority of users can operate them with as few as one recording channel, minimal setup, calibration, and training.

Virtual reality (VR) has been validated as effective, safe, affordable and motivating feedback approach for BCI systems (Lotte 2008, Leeb 2009). Augmented reality (AR) enhances the physical world by superimposing informative, context sensitive computer generated content (Schmalstieg et al. 2002). Navarro (2004) proposed the use of wearable BCI systems in AR. Recent studies like (Bell et al. 2008) and (Kansaku et al. 2010) report on BCI driven desktop-based real-world interaction using camera-equipped robotic agents.

In this paper, we extend our work on SSVEP BCIs and desktop-based VR (Faller et al. 2010a,b) to SSVEP BCI based operation in immersive VR and AR environments. This investigation is very interesting since AR user interfaces could compensate for the low bandwidth of BCIs by offering a more direct, friendly and intuitive interface to the physical world, hence facilitating a more seamless interaction.

# 2. Materials and methods

*2.1. Subjects*

Three male subjects (aged 26-27; two SSVEP experienced, one BCI naive) free of neurological disorders or medication that might adversely affect the EEG, voluntarily participated in the study. All subjects gave written informed consent prior to the experiment and were reimbursed with 7.50 € per hour for their time. Nature and purpose of the experiment were explained in personal communication supported by written instructions.

*2.2. Signal acquisition and processing*

Signals were derived according to the 10-20 system (Jasper 1958) using three sintered AgCl electrodes in a bipolar setup, 2.5 cm anterior and posterior to O1 with the ground-electrode placed at Fpz. We kept the impedances below 5 kΩ. The data was recorded using a biosignal amplifier (g.tec, Guger Technologies, Graz, Austria), a data acquisition card (NI-6031E, National Instruments Corporation, Austin, Texas) and a standard Windows XP PC (Microsoft Corporation, Redmond, Washington). We applied a bandfilter between 0.5 and 100 Hz, a notch filter at 50 Hz and sampled at 256 Hz. The data was processed in real-time using rtsBCI (Schlögl & Brunner 2008) and the classification method harmonic sum decision (HSD, see Müller-Putz et al. 2008).

*2.3. Experimental setup and paradigm*

Feedback and SSVEP stimuli were rendered on a dedicated Windows XP PC (Intel Core i5 750, 4096 MB RAM, NVidia GeForce GTX 260) and presented with a head-mounted display (HMD; V8 Virtual Research Systems, Aptos, California). The real-world video image for the AR scenario was acquired online using a USB camera (Logitech Webcam Pro 9000, Logitech Inc., Fremont, California) mounted on top of the HMD. All feedback was generated using the mixed reality framework Studierstube (Schmalstieg et al. 2002) along with ARToolKitPlus (Wagner & Schmalstieg 2007) for the AR scenario.

The task for the VR and AR condition was identical: The subjects had to wait for 30 s, then activate the navigation stimuli, guide the avatar through the slalom to the second gray waypoint, where they had to deactivate the navigation stimuli again. The run ended after 30 more seconds (see Figure 1.A). Maximum time to task completion was ten minutes. There were two runs for each condition. All stimuli were represented as quadratic planes steadily oscillating between red and black at 12, 15, 20 and 8 Hz. The three stimuli next to the avatar were associated with three navigation commands (see Figure 1.B). The fourth stimulus was statically placed at the top right, and was used for switching the navigation stimuli and the associated BCI detectors on and off to allow for more stability during no-control periods (similar to Cheng et al. 2002). The camera was fixed in the VR scenario, and the angle was similar to that in the AR condition where the camera was mounted on the HMD. The subjects answered short questionnaires after finishing a condition.

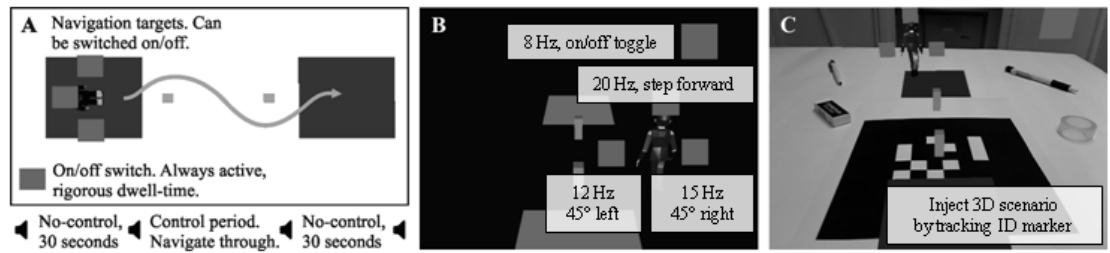

**Figure 1** The speaker symbols in panel A indicate audible cues that notified the subject of the state transitions in the task. The screenshots in panel B and C are from actual online runs. The associated commands and frequencies annotated in the screenshot of the VR condition in panel B are identical for both conditions. The 3D graphics in the AR condition seen in panel C are tracked to the fiducial marker in the screenshot.

*2.4. Classification*

According to the HSD method, one class is selected as soon as the sum of the responses at the base frequency along with the second and third harmonic components exceeds the sums for all other responses throughout a dwell-time of 1 s for the navigation stimuli and 1.5 s for the on/off stimulus. The responses for the target frequencies were normalized using data from 1 minute calibration measurements before the VR and AR scenario. A 3 s refractory period followed every activation.

*2.5. Performance evaluation*

We report intentional (task-conform) interactions in control state as true positives $TP_C$ per minute and unintentional interactions as false positives $FP_C$ and $FP_{NC}$ per minute for control and no-control period respectively. We calculate the positive predictive value (PPV or precision, Altman & Bland 1994) over control and no-control periods of the runs ($PPV=TP_C/(TP_C+FP_C+FP_{NC})$), plot the navigation trajectories and list the time to task completion.

# 3. Results

Figure 2 shows example trajectories for the three subjects in the VR and AR condition.

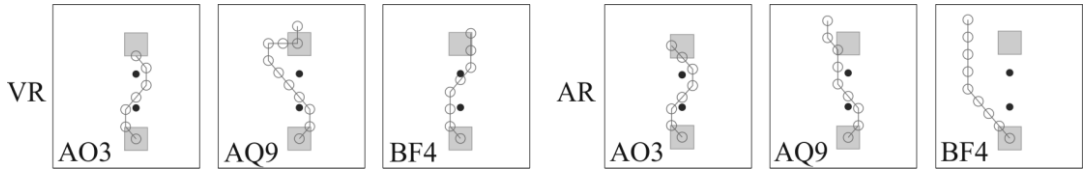

**Figure 2** The two filled dots in these example trajectories for the three subjects in the two conditions mark the poles of the slalom.

Table 1 summarizes the most characteristic performance measures as described in 2.5. The values represent the averages over the performed runs.

**Table 1** Evaluation results for the VR and AR condition.

|      | PPV (%) | | $TP_C$ (min$^{-1}$) | | $FP_C$ (min$^{-1}$) | | $FP_{NC}$ (min$^{-1}$) | | Time (s) | |
|------|------|------|------|------|------|------|------|------|------|------|
|      | VR   | AR   | VR   | AR   | VR   | AR   | VR   | AR   | VR   | AR   |
| AO3  | 87.1 | 77.5 | 5.5  | 4.7  | 0.6  | 0.9  | 0.5  | 1.5  | 148  | 199  |
| AQ9  | 78.7 | 77.1 | 4.6  | 4.1  | 1.1  | 0.8  | 0.5  | 1.5  | 240  | 200  |
| BF4  | 70.4 | ---  | 4.2  | ---  | 1.3  | ---  | 2.0  | ---  | 270  | ---  |
| Mean | 78.7 | 77.3 | 4.8  | 4.4  | 1.0  | 0.8  | 1.0  | 1.5  | 219  | 199  |
| Std  | 8.4  | 0.0  | 0.6  | 0.4  | 0.4  | 0.1  | 0.9  | 0.0  | 63   | 0    |

# 4. Discussion

Both the immersive VR and AR scenarios proved to be effective setups for feedback and dynamic SSVEP stimulus presentation. To our knowledge, this is the first work to report on a SSVEP BCI that operates using target stimuli which are integrated in immersive VR and AR. The experienced subjects (AO3 and AQ9) finished all runs in both conditions successfully whereas the naive subject BF4 achieved moderate control only in the first run of the VR condition. The observed decrease in the PPV and increase in the number of FPs in non-control state from the VR to the AR condition goes in line with the subjects' report in the questionnaire that they found the AR condition slightly more difficult. Reasons may include the higher dynamic of the scenery, the slight changes in the point of view when moving the head or the natural, maybe distracting real-world environment. See-through HMDs and background sensitive adjustment of contrast or color of the stimuli could be ways to optimize this. These questions require further investigation. AO3 and AQ9 found both scenarios very motivating, while BF4 felt neutral in this concern. All subjects felt neutral as far as annoyance of the oscillating stimuli was concerned. The fact that all of the subjects feel neutral or even positive about using a system like this in a real-world setting (user acceptance) support the argument, that SSVEP BCIs with AR user interfaces can become viable communication devices.

# 4. Conclusion

AR can improve real-world practicality and usability of BCI systems by compensating for some of their traditional shortcomings such as the low bandwidth by offering a richer, more direct, and intuitive interface. This allows for a more goal-directed and seamless real-world interaction. AR user interfaces may combine particularly well with SSVEP based BCIs, since an arbitrary number of stimuli can be spatially associated to distinct points of interest in the physical world. These may be abstract or may overlap physical objects such as devices, people or controls. This can be an elegant and intuitive way of presenting the user with all possible interaction options. Also, SSVEP-based BCIs have been shown to be especially effective for selection tasks (Cheng et al. 2002, Gao et al. 2003). These systems could provide patients with a higher degree of self autonomy and functional independence by introducing more intuitive and effective smart home control. Also, the combination of AR and BCI technology can introduce a valuable, additional communication or control channel for user groups that require or benefit from hands free operation (e.g. due to temporary situational disability) like pilots, astronauts, drivers or office workers.

## Acknowledgement

This work was supported by the EU research project Brainable ICT-2009-4-247447.